\documentclass[aps,showpacs,twocolumn,amssymb]{revtex4}
\usepackage{amssymb}
\usepackage{graphicx}
\usepackage{color}
\usepackage{textcomp}
\usepackage{ulem}
\begin{document}

\title{Chiral Symmetry Restoration and Quark Deconfinement beyond Mean Field\\ in a Magnetized PNJL Model}
\author{Shijun Mao}
\affiliation{School of Science, Xi'an Jiaotong University, Xi'an, Shaanxi 710049, China}

\begin{abstract}
We study chiral symmetry restoration and quark deconfinement beyond mean field approximation in a magnetized PNJL model. The feedback from mesons to quarks modifies the quark coupling constant and Polyakov potential. As a result, the separate critical temperatures for the two phase transitions at mean field level coincide and the magnetic catalysis becomes inverse magnetic catalysis, when the meson contribution is included.
\end{abstract}

\date{\today}
\maketitle

Chiral symmetry restoration and quark deconfinement are the two most important Quantum Chromodynamics (QCD) phase transitions at finite temperature and baryon density. Motivated by the strong magnetic field in the core of compact stars and in the initial stage of relativistic heavy ion collisions, the study on QCD phase structure is recently extended to including external electromagnetic fields, see reviews~\cite{review1,review2,review3,review4,review5}. Due to the Fermion dimension reduction, a magnetic catalysis effect on chiral symmetry breaking is expected in both vacuum and finite temperature in almost all model calculations at mean field level, see for instance \cite{mc1,mc2,mc3}. However, from the recent lattice QCD simulations with a physical pion mass, while the chiral condensate is enhanced in vacuum, the critical temperature of the phase transition drops down with increasing magnetic field~\cite{lattice1,lattice2,lattice3,lattice4,lattice5}. To understand this inverse magnetic catalysis at high temperature, many scenarios are proposed~\cite{fukushima,mao,kamikado,bf1,bf12,bf13,bf2,bf3,bf4,bf5,bf6,bf7,bf8,bf9,bf11,db1,db2,db3,db5,db6,pnjl1,pnjl2,pnjl3,pnjl4,pqm}, such as magnetic inhibition of mesons, mass gap in large $N_c$ limit, sphalerons, gluon screening effect, and weakening of strong coupling.

The lattice simulations on the Polyakov loop which is considered as the order parameter of deconfinement phase transition support the inverse magnetic catalysis, namely a decreasing critical temperature as the magnetic field grows~\cite{lattice4,lattice5}. The deconfinement phase transition in an external magnetic field is also widely investigated in effective models, such as the MIT bag model~\cite{mit1,mit2}, Polyakov-extended Nambu--Jona-Lasinio Model (PNJL)~\cite{pnjl1,pnjl2,pnjl3,pnjl4} and Polyakov-extended Quark-Meson model~\cite{pqm}. However, how to understand the inverse magnetic catalysis is still an open question. In this paper, we investigate the magnetic field effect on chiral symmetry restoration and quark deconfinement in a PNJL model~\cite{pnjl5,pnjl6,pnjl7,pnjl8,pnjl9,pnjl10,pnjl11,pnjl12} beyond mean field approximation. By extending the method of including feedback from mesons to quarks in the NJL model~\cite{mao,zhuang} to the PNJL model, we focus on the behavior of the chiral condensate and Polyakov loop in finite temperature and magnetic field.

The two-flavor PNJL model in external electromagnetic and gluon fields is defined through the Lagrangian density~\cite{pnjl5,pnjl6,pnjl7,pnjl8,pnjl9,pnjl10,pnjl11,pnjl12} in chiral limit,
\begin{equation}
\label{pnjl}
{\cal L} = i\bar\psi\gamma_\mu D^\mu\psi + {G\over 2}\left[\left(\bar\psi\psi\right)^2 + \left(\bar\psi i \gamma_5 {\bf \tau}\psi\right)^2 \right]-{\cal U}(\Phi,\bar\Phi)
\end{equation}
with the Polyakov potential describing deconfinement at finite temperature,
\begin{equation}
\label{polyakov}
{{\cal U}\over T^4} = -{b_2\over 2} \bar\Phi\Phi -{b_3\over 6}\left({\bar\Phi}^3+\Phi^3\right)+{b_4\over 4}\left(\bar\Phi\Phi\right)^2,
\end{equation}
where the coefficient $b_2(t)=a_0+a_1 t+a_2 t^2+a_3 t^3$ with $t=T_0/T$ is temperature dependent, and the other coefficients $b_3$ and $b_4$ are constants. The covariant derivative $D^\mu=\partial^\mu+i Q A^\mu-i {\cal A}^\mu$ couples quarks to the two external fields, the magnetic field ${\bf B}=\nabla\times{\bf A}$ and the temporal gluon field  ${\cal A}^\mu=\delta^\mu_0 {\cal A}^0$ with ${\cal A}^0=g{\cal A}^0_a \lambda_a/2=-i{\cal A}_4$ in Euclidean space. The gauge coupling $g$ is combined with the SU(3) gauge field ${\cal A}^0_a(x)$ to define ${\cal A}^\mu(x)$, $\lambda_a$ are the Gell-Mann matrices in color space, and $Q=diag(Q_u, Q_d)=diag(2e/3,-e/3)$ is the quark charge matrix in flavor space. To simplify calculations, we assume magnetic field ${\bf B}=(0, 0, B)$ along the $z$-axis. For the chiral section in the Lagrangian, $G$ is the coupling constant in the scalar and pseudo-scalar channels. The order parameter to describe chiral phase transition is the chiral condensate $\langle\bar\psi\psi\rangle$ or the dynamical quark mass $m=-G \langle \bar\psi\psi\rangle$. In chiral limit, the SU$_L(2) \otimes$ SU$_R(2)$ symmetry of the system is broken down to U(1)$_L\otimes$ U(1)$_R$ by the magnetic field ${\bf B}$, and the number of Goldstone modes is reduced from 3 ($\pi_0$ and $\pi_\pm$) to 1 ($\pi_0$). The Polyakov potential ${\cal U}(\Phi,{\bar \Phi})$ is related to the Z(3) center symmetry and simulates the deconfinement in terms of the trace of the Polyakov loop $\Phi=\left({\text {Tr}}_c L \right)/N_c$, where $L({\bf x})={\cal P} \text {exp}[i \int^\beta_0 d \tau A_4({\bf x},\tau)]= \text {exp}[i \beta A_4 ]$ with $\beta=1/T$ is a matrix defined in color space~\cite{pnjl6}. $\Phi$ can be considered as the order parameter to describe the deconfinement process, since it satisfies $\Phi \rightarrow 0$ in confined phase at low temperature and $\Phi \rightarrow 1$ in deconfined phase at high temperature~\cite{pnjl5,pnjl6,pnjl7,pnjl8,pnjl9,pnjl10,pnjl11,pnjl12}. Note that there is $\Phi=\bar\Phi$ at vanishing baryon density.

In mean field approximation, the thermodynamic potential of the quark system contains the mean field part and quark part~\cite{pnjl1,pnjl2,pnjl3,pnjl4,pnjl5,pnjl6,pnjl7,pnjl8,pnjl9,pnjl10,pnjl11,pnjl12},
\begin{eqnarray}
\label{omega1}
\Omega_{mf} &=&{\cal U}(\Phi)+ \frac{m^2}{2 G}+\Omega_q,\\
\Omega_q &=& - \sum_{f,n}\alpha_n \int \frac{d p_z}{2\pi} \frac{|Q_f B|}{2\pi} \big[3E_f\nonumber\\
&+& 2T\ln\left(1+3\Phi e^{-\beta E_f}+3{ \Phi}e^{-2\beta E_f}+e^{-3\beta E_f}\right)\big]\nonumber
\end{eqnarray}
with spin factor $\alpha_n=2-\delta_{n0}$ and quark energy $E_f=\sqrt{p^2_z+2 n |Q_f B|+m^2}$ for flavor $f$ and Landau level $n$. The ground state is determined by minimizing the thermodynamic potential, $\partial\Omega_{mf}/\partial m=0$ and $\partial\Omega_{mf}/\partial \Phi=0$, which leads to the two coupled gap equations for the two order parameters $m$ and $\Phi$,
\begin{eqnarray}
\label{gap1}
&& m\left( \frac{1}{2G}+\frac{\partial \Omega_q}{\partial m^2}\right)=0,\nonumber \\
&& \frac{\partial {\cal U}}{\partial \Phi}+\frac{\partial \Omega_q}{\partial \Phi}=0.
\end{eqnarray}
Solving the two mean field gap equations, the chiral symmetry restoration and quark deconfinement happen around $T_c \simeq 220$ MeV in case of vanishing magnetic field. When the field is turned on, there is magnetic catalysis for both the chiral symmetry restoration and quark deconfinement phase transitions, see for instance \cite{pnjl1,pnjl2,pnjl3,pnjl4}.

In NJL models, mesons are treated as quantum fluctuations above the mean field and constructed through random phase approximation (RPA)~\cite{njl1,njl2,njl3,njl4,njl5}. There are four mesons in the case of two flavors, the isospin singlet $\sigma$ and triplet $\pi_0$ and $\pi_\pm$, corresponding to the scalar and pseudoscalar channels in the Lagrangian. Taking into account the fact that the difference between NJL and PNJL models is the background field $\Phi$, while it changes the medium properties, it does not dynamically affect the interaction between quarks. Therefore, the meson propagator in the PNJL model can still be written as~\cite{njl6,fu}
\begin{eqnarray}
D_{M}(k^2_0,{\bf k}^2)=\frac{G}{1-G\Pi_M(k^2_0,{\bf k}^2)}
\end{eqnarray}
with the quark bubble or meson polarization function $\Pi_M(k_0^2,{\bf k}^2)$. The meson pole mass $m_M$ and coupling constant $g_{Mq\bar q}$ are defined at the pole of the meson propagator at zero momentum~\cite{mao,ritus1,ritus2},
\begin{eqnarray}
\label{pole}
&& 1-G\Pi_M(m_M^2,{\bf 0})=0,\\
&& \left(g_{Mq\bar q}^\mu\right)^2=\left[g^{\mu\mu}{\partial\Pi_M(k_0^2,{\bf k}^2)\over \partial k_\mu^2}\Big|_{(k_0^2,{\bf k}^2)=(m_M^2,{\bf 0})}\right]^{-1}\nonumber
\end{eqnarray}
with space-time metric $g^{\mu\nu}=diag(1,-1,-1,-1)$. Note that the magnetic field reduces the Lorentz group SO(1,3) of space-time transformation to its subgroup SO(1,1) in the direction of the field, which leads to an anisotropic coupling constant $g^\mu_{Mq\bar q}$ with elements $g^1_{Mq\bar q}=g^2_{Mq\bar q}\neq g^0_{Mq\bar q}=g^3_{Mq\bar q}$.

After a straightforward calculation, the polarization functions for $\pi_0$ and $\sigma$ at the pole can be simplified as
\begin{eqnarray}
\label{bubble2}
\Pi_M(k_0^2,0) &=& 3\sum_{f,n}\alpha_n \left|\frac{Q_f B}{2\pi}\right| \int \frac{d p_z}{2\pi}\\
&\times& \left[\frac{E_f^2-\epsilon_M^2/4}{E_f^2-k_0^2/4} \frac{1-2f_\Phi(E_f)}{E_f}\right]\nonumber
\end{eqnarray}
with meson factors $\epsilon_{\pi_0}=0$ and $\epsilon_\sigma=2m$ and the modified quark distribution function
\begin{equation}
f_\Phi(E_f)=\frac{\Phi e^{-\beta E_f}+ 2\Phi e^{-2\beta E_f}+ e^{-3\beta E_f}}{1+3\Phi e^{-\beta E_f}+ 3\Phi e^{-2\beta E_f}+ e^{-3\beta E_f}}.
\end{equation}
For $\Phi=1$, $f_\Phi(E_f)$ is reduced to the Fermi-Dirac distribution function, and the quark thermodynamical potential, quark mass gap equation and the meson polarization function will recover the result in the NJL model~\cite{mao}.

By comparing the gap equation for quark mass $m$ at mean field level with the pole equation for neutral meson mass $m_M$ at RPA level, there exist simple relations in the chiral symmetry breaking phase,
\begin{equation}
\label{mass}
m\neq 0,\ \ \ \ \ m_{\pi_0} = 0,\ \ \ \ \ m_\sigma = 2m.
\end{equation}
This confirms that $\pi_0$ is the Goldstone mode corresponding to the spontaneous chiral symmetry breaking. As a consequence, neutral pions and quarks dominate respectively the thermodynamics of the system in chiral symmetry breaking phase at low temperature and chiral symmetry restoration phase at high temperature. Around the critical point, the Higgs mode $\sigma$ will also play an important role. To avoid the complicated calculation for charged pions which become heavier in magnetic field, we consider in the following only neutral mesons $\pi_0$ and $\sigma$, which is a good approximation in strong magnetic field (in week magnetic field the mass difference between neutral and charged pions is not significant).

Including meson degrees of freedom in the model, the thermodynamic potential of the quark-meson plasma can be generally written as
\begin{equation}
\label{omega2}
\Omega={\cal U}(\Phi)+{m^2\over 2G}+\Omega_q+\sum_M\Omega_M,
\end{equation}
where $\Omega_M$ is the meson thermodynamic potential. In pole approximation, mesons are treated as quasi-particles and $\Omega_M$ can be simply expressed as
\begin{equation}
\label{omegam1}
\Omega_M = \int \frac{d^3 {\bf k}}{(2\pi)^3} \left[\frac{E_M}{2} +T \ln\left(1-e^{-E_M/T}\right)\right]
\label{omeson}
\end{equation}
with the meson energy $E_M = \sqrt{m_M^2+k_3^2+v^2_\perp (k_1^2+k_2^2)}$, where $v_\perp =g^3_{Mq\bar q}/g^1_{Mq\bar q}$ is called the meson transverse velocity~\cite{mao,ritus1,ritus2}. Due to the anisotropy introduced by the external magnetic field, there is $v_\perp \neq 1$ in general case.

With the total thermodynamic potential (\ref{omega2}), the order parameters $m$ and $\Phi$ beyond mean field approximation should correspond to the minimum of the new potential, $\partial\Omega/\partial m=0$ and $\partial\Omega/\partial\Phi=0$, namely
\begin{eqnarray}
\label{gap2}
&& m\left( \frac{1}{2G}+\frac{\partial \Omega_q}{\partial m^2}+\sum_M\frac{\partial \Omega_M}{\partial m^2}\right) =0,\nonumber\\
&& \frac{\partial {\cal U}}{\partial \Phi}+\frac{\partial \Omega_q}{\partial \Phi}+\sum_M\frac{\partial \Omega_M}{\partial \Phi}=0.
\label{newgap}
\end{eqnarray}
Obviously, the new order parameters $m$ and $\Phi$ determined by (\ref{gap2}) are different from the mean field ones determined by (\ref{gap1}), and the difference comes from the thermodynamic contribution from mesons.

Taking into account the fact that in quark models quarks are elementary constituents and mesons are quantum fluctuations, the meson induced correction to the order parameters should be small, $|m-m_{mf}|/m_{mf}\ll 1$ and $|\Phi-\Phi_{mf}|/\Phi_{mf}\ll 1$. Therefore, we can expand the meson thermodynamics around the mean field~\cite{zhuang,mao} and keep only the first order derivatives to simplify the numerical calculation,
\begin{eqnarray}
\label{omeham2}
\Omega_M &=& \sum_{i,j}{1\over i!j!} {\partial^{i+j}\Omega_M\over \partial (m^2)^i\partial \Phi^j}\Big|_{mf}\left(m^2-m_{mf}^2\right)^i\left(\Phi-\Phi_{mf}\right)^j\nonumber\\
&\simeq& \Omega_M|_{mf}+{\partial\Omega_M\over \partial m^2}\Big|_{mf}\left(m^2-m_{mf}^2\right)\nonumber\\
&+&{\partial\Omega_M\over \partial \Phi}\Big|_{mf}\left(\Phi-\Phi_{mf}\right).
\end{eqnarray}
Under this approximation, we can rewrite the new gap equations (\ref{gap2}) in the same form as the mean field ones (\ref{gap1}),
\begin{eqnarray}
\label{gap3}
&& m\left( \frac{1}{2G'}+\frac{\partial \Omega_q}{\partial m^2}\right) =0,\nonumber\\
&&\frac{\partial {\cal U}'}{\partial \Phi}+\frac{\partial \Omega_q}{\partial \Phi}=0.
\end{eqnarray}
The meson contribution here is fully absorbed into the effective coupling constant $G'$ and the effective Polyakov potential ${\cal U}'$ defined by
\begin{eqnarray}
\label{newg}
{1\over 2G'} &=& {1\over 2G}+\sum_M{\partial\Omega_M\over\partial m^2}\Big|_{mf},\nonumber\\
{\cal U}' &=& {\cal U}+\sum_M{\partial\Omega_M\over\partial\Phi}\Big|_{mf}\Phi,
\end{eqnarray}
where we have neglected the $\Phi$-independent term $\sum_M[\Omega_M|_{mf}+\partial\Omega_M/\partial m^2|_{mf}(m^2-m_{mf}^2)-\partial\Omega_M/\partial\Phi|_{mf}\Phi_{mf}]$ in the effective potential ${\cal U}'$, since it does not contribute to the second gap equation for $\Phi$. Different from the original coupling $G$ which is a constant, the effective coupling $G'(T,B)$ is defined in hot medium and magnetic field. The effective Polyakov potential ${\cal U}'$ depends also the magnetic field, in addition to the temperature dependence in the original potential ${\cal U}$. Note that, the quantum fluctuations affect not only the chiral phase transition through the effective coupling but also the quark deconfinement through the linear term in the new Polyakov potential ${\cal U}'$.

\begin{figure}[hb]
\centering
\includegraphics[width=7.5cm]{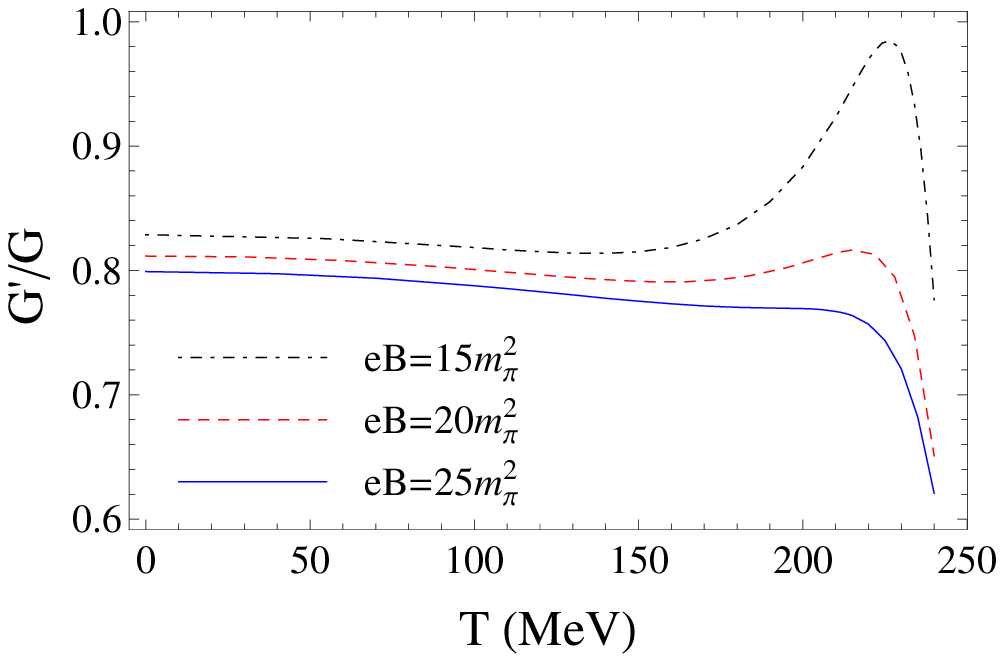}
\includegraphics[width=7.5cm]{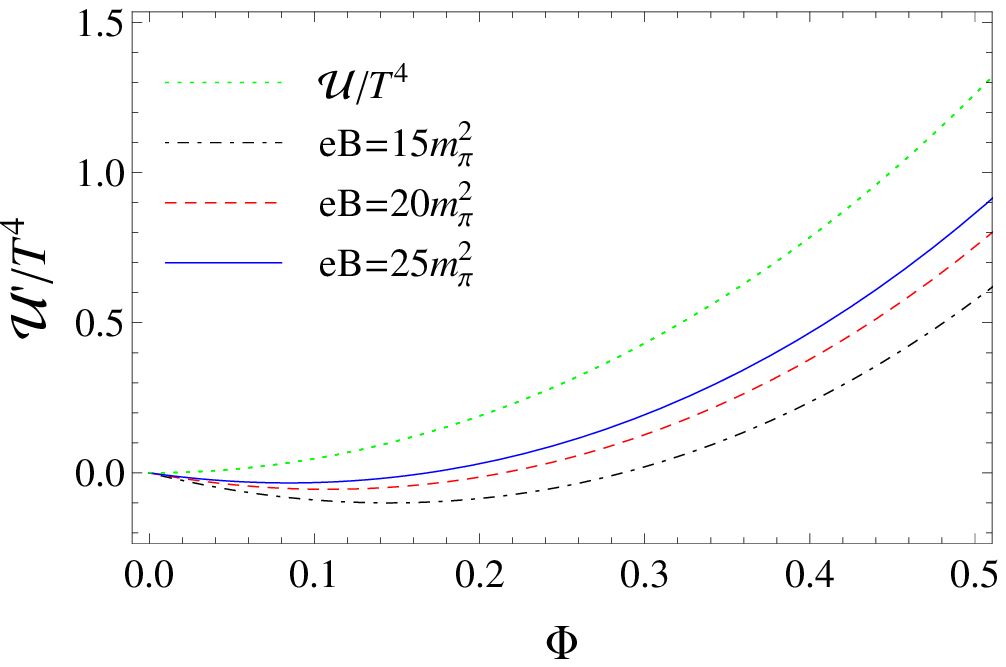}
\caption{The dimensionless coupling constant $G'/G$ as a function of temperature and Polyakov potential ${\cal U}'/T^4$ (fixing $T=200$ MeV) as a function of Polyakov loop at different magnetic field. }
\label{fig1}
\end{figure}
Because of the contact interaction among quarks, NJL models are non-renormalizable, and it is necessary to introduce a regularization scheme to remove the ultraviolet divergence in momentum integrations. To guarantee the law of causality in magnetic field (the transverse velocity $v_\perp < 1$ for the Goldstone mode), we take a covariant Pauli-Villars regularization~\cite{mao}. To do numerical calculations at finite temperature and magnetic field, we need to fix the model parameters in vacuum with $T=B=0$. In chiral limit there are two parameters, the quark coupling constant $G$ and momentum cutoff $\Lambda$. By fitting the pion decay constant $f_\pi=93$ MeV and chiral condensate $\langle \bar \psi \psi \rangle =(-250\ \text{MeV})^3$ in vacuum, the two parameters are fixed to be $G=7.79$ GeV$^{-2}$ and $\Lambda=1127$ MeV in Pauli-Villars scheme with number of regulated quark masses $N=3$. On the Polyakov potential, its temperature dependence is from the lattice simulation, and the parameters are chosen as~\cite{pnjl6} $a_0=6.75$, $a_1=-1.95$, $a_2=2.625$, $a_3=-7.44$, $b_3=0.75$, $b_4=7.5$ and $T_0=270$ MeV. Considering the fact that charged pions may have sizeable contribution to the thermodynamics of the system in weak magnetic field, we have taken a safe region $eB\ge 15 m_\pi^2$ in our numerical calculations.

We first solve the coupled gap equations (\ref{gap1}) to obtain the mean field quark mass $m_{mf}$ and Polyakov loop $\Phi_{mf}$, and then substitute them into the pole equation (\ref{pole}) to calculate the meson mass $m_M$ and meson coupling constant $g^{\mu}_{Mq\bar q}$. For the Goldstone mode $\pi_0$, its longitudinal velocity is exactly the speed of light, $v_{||}=1$, but its transverse velocity $v_{\perp}$ is always less than the speed of light at any temperature and magnetic field, satisfying the law of causality. With the known $m_{mf}$ and $\Phi_{mf}$ at mean field level and $m_M$ and $v_\perp$ at RPA level, we further calculate the effective coupling constant $G'$ and Polyakov potential ${\cal U}'$ through the definition (\ref{newg}).

Fig.\ref{fig1} shows the dimensionless coupling constant $G'/G$ as a function of temperature and Polyakov potential ${\cal U}'/T^4$ as a function of Polyakov loop. In mean field approximation, we have $G'/G=1$ and ${\cal U}'$ is reduced to ${\cal U}$, which are independent of the magnetic field. The quantum fluctuations weaken the coupling among quarks, and there is always $G'<G$ at any $T$ and $B$. At fixed $T$, the coupling drops down with increasing $B$, indicating the magnetic inhabition of mesons~\cite{fukushima}. The contribution from mesons also modifies the Polyakov potential. In comparison with the potential ${\cal U}$ in mean field approximation, the location of the minimal potential ${\cal U}'$ shifts towards a larger $\Phi$, which means that the contribution from mesons accelerates the deconfinement phase transition.

\begin{figure}[hb]
\centering
\includegraphics[width=8.5cm]{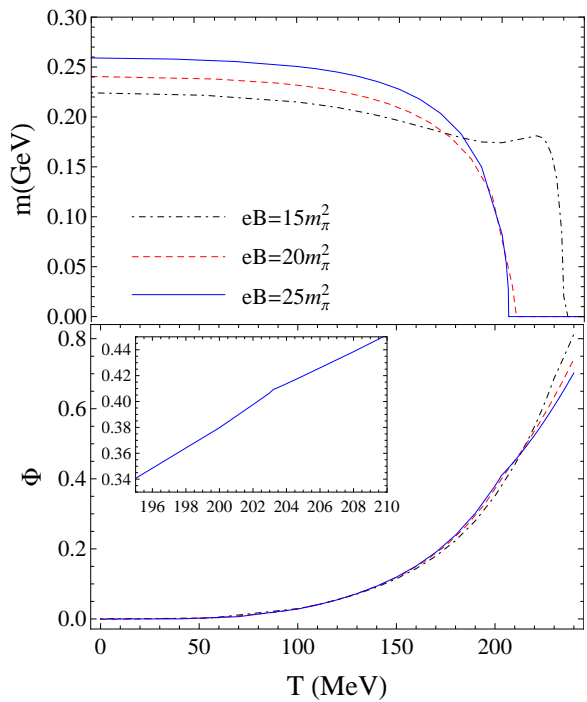}
\caption{The quark mass $m$ and Polyakov loop $\Phi$ as functions of temperature at different magnetic field. }
\label{fig2}
\end{figure}
With the known effective coupling $G'$ and effective potential ${\cal U}'$, we now calculate the quark mass $m$ and Polyakov loop $\Phi$ beyond mean field approximation through the coupled gap equations (\ref{gap3}). Fig.\ref{fig2} shows $m$ and $\Phi$ as functions of temperature at different magnetic field. While $m$ in vacuum goes up with magnetic field, the critical temperature $T_\chi$ for chiral phase transition decreases with magnetic field, indicating magnetic catalysis at low temperature and inverse magnetic catalysis at high temperature, which are in good agreement with the lattice QCD simulations~\cite{lattice1,lattice2,lattice3,lattice4,lattice5}. Different from the behavior of the quark mass $m$, the Polyakov loop $\Phi$ is almost independent of the magnetic field at low temperature, it is very close to the vacuum value (zero). When temperature approaches to the critical point $T_\chi$ of chiral phase transition, $\Phi$ starts to increase clearly and more and more fast. Going beyond the critical point, chiral symmetry is restored with $m=0$, and the two gap equations (\ref{gap3}) are decoupled. In this case, the meson contribution to the deconfinement is reflected only in the linear term in the potential, which leads to a slower increase of the order parameter $\Phi$. The inflection located at $T_\chi$ becomes more clear with increasing magnetic field, see the small figure for $eB=25 m_\pi^2$ in the lower panel of Fig.\ref{fig2}.

\begin{figure}[hbt]
\centering
\includegraphics[width=8cm]{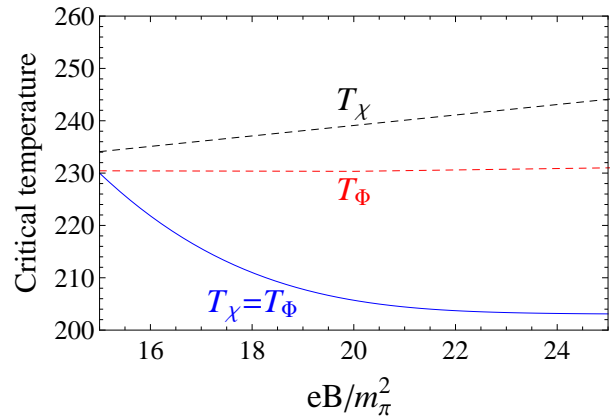}
\caption{ The critical temperatures $T_\chi$ for chiral symmetry restoration and $T_\Phi$ for quark deconfinement as functions of magnetic field in (dashed lines) and beyond (solid line) mean field approximation. }
\label{fig3}
\end{figure}
To explicitly demonstrate the magnetic field effect on chiral symmetry restoration and quark deconfinement, we plot the chiral restoration temperatures $T_\chi$ and deconfinement temperature $T_\Phi$ as functions of scaled magnetic field in Fig.\ref{fig3}. In chiral limit, $T_\chi(B)$ is well defined through the condition $m(T_\chi,B)=0$ and the phase transition is of second order. For the deconfinement which is a crossover in finite temperature and magnetic field, $T_\Phi(B)$ is usually defined as the temperature where the Polyakov loop has the maximum change~\cite{pnjl1,pnjl2,pnjl3,pnjl4,pnjl5,pnjl6,pnjl7,pnjl8,pnjl9,pnjl10,pnjl11,pnjl12}. In mean field approximation, there are respectively strong and weak magnetic catalysis effect on the chiral symmetry restoration and quark deconfinement, see the dashed lines in Fig.\ref{fig3}. When the feedback from mesons to quarks is included, however, the behavior of the two temperatures in magnetic field is significantly changed. $T_\chi$ drops down with increasing magnetic field, clearly shown in Fig.\ref{fig2}. Since the maximum change of the Polyakov loop is located at $T_\chi$, as shown in Fig.\ref{fig2}, the deconfinement temperature $T_\Phi$ coincides with $T_\chi$. Therefore, the meson contribution removes the difference between the chiral restoration and deconfinement temperatures, and there is inverse magnetic catalysis effect on both chiral symmetry restoration and quark deconfinement, which are consistent with the lattice QCD simulations~\cite{lattice1,lattice2,lattice3,lattice4,lattice5}.

In summary, the magnetic field effect on chiral symmetry restoration and quark deconfinement is investigated in the frame of a two-flavor PNJL model beyond mean field approximation. The feedback from mesons to quarks is reflected in the modification of the quark coupling constant and Polyakov potential, which accelerates the chiral symmetry restoration and quark deconfinement. By going from mean field approximation to including meson contributions, the difference between the chiral restoration temperature and deconfinement temperature is removed, and the magnetic catalysis effect on the two phase transitions becomes inverse magnetic catalysis.

\noindent {\bf Acknowledgement:} The work is supported by the NSFC grant 11405122, 11775165.


\begin{thebibliography}{99}

\bibitem{review1} F.Preis, A.Rebhan and A.Schmitt, Lect. Notes Phys. {\bf 871}, 51(2013).
\bibitem{review2} R.Gatto and M.Ruggieri, Lect. Notes Phys. {\bf 871}, 87(2013).
\bibitem{review3} M.D'Elia, Lect. Notes Phys. {\bf 871}, 181(2013).
\bibitem{review4} V.A.Miransky and I.A.Shovkovy, Phys. Rep. {\bf 576}, 1(2015).
\bibitem{review5} J.O.Anderson and W.R.Naylor, Rev. Mod. Phys. {\bf 88}, 025001(2016).
\bibitem{mc1} S.P.Klevansky and R.H.Lemmer, Phys. Rev. {\bf D39}, 3478 (1989).
\bibitem{mc2} K.G.Klimenko, Theor. Math. Phys. {\bf 89}, 1161 (1992).
\bibitem{mc3} V.P.Gusynin, V.A.Miransky and I.A.Shovkovy, Nucl. Phys. {\bf B462}, 249(1996).
\bibitem{lattice1} G.S.Bali {\it et al.}, JHEP {\bf 1202}, 044(2012).
\bibitem{lattice2} G.S.Bali {\it et al.}, Phys. Rev. {\bf D86}, 071502(2012).
\bibitem{lattice3} G.S.Bali {\it et al.}, JHEP {\bf 04}, 130(2013).
\bibitem{lattice4} V.Bornyakov {\it et al.}, Phys. Rev. {\bf D90}, 034501(2014).
\bibitem{lattice5} G.Endrodi, JHEP{\bf 07}, 173(2015).

\bibitem{fukushima} K.Fukushima and Y.Hidaka, Phys. Rev. Lett {\bf 110}, 031601(2013).
\bibitem{mao} S.J.Mao, Phys. Lett. {\bf B758}, 195(2016), Phys. Rev. {\bf D94}, 036007(2016).
\bibitem{kamikado}  K.Kamikado and T.Kanazawa, JHEP {\bf 03}, 009(2014).
\bibitem{bf1} J.Y.Chao, P.C.Chu and M.Huang, Phys. Rev. {\bf D88}, 054009(2013).
\bibitem{bf12} F.Bruckmann, G.Endrodi and T.G.Kovacs, arXiv: 1311.3178.
\bibitem{bf13} J.Braun, W.A.Mian and S.Rechenberger, Phys. Lett. {\bf B755}, 265(2016).
\bibitem{bf2} N.Mueller and J.M.Pawlowski,  Phys. Rev. {\bf D91}, 116010(2015).
\bibitem{bf3} T.Kojo and N.Su, Phys. Lett. {\bf B720}, 192(2013).
\bibitem{bf4} F.Bruckmann, G.Endrodi and T.G.Kovacs, JHEP{\bf 1304}, 112(2013).
\bibitem{bf5} A.Ayala {\it et al.}, Phys. Rev. {\bf D92}, 096011(2015).
\bibitem{bf6} A.Ayala {it et al.},  Phys. Rev. {\bf D90}, 036001(2014).
\bibitem{bf7} A.Ayala {\it et al.}, Phys. Rev. {\bf D89}, 116017(2014).
\bibitem{bf8} R.L.S.Farias {\it et al.}, Phys. Rev. {\bf C90}, 025203(2014).
\bibitem{bf9} M.Ferreira {\it et al.}, Phys. Rev. {\bf D89}, 116011(2014).
\bibitem{bf11} F.Preis, A.Rebhan and A.Schmitt, JHEP {\bf 1103}, 033(2011).
\bibitem{db1} E.S.Fraga and A.J.Mizher, Phys. Rev. {\bf D78}, 025016(2008), Nucl. Phys. {\bf A820}, 103C(2009).
\bibitem{db2} K.Fukushima, M.Ruggieri and R.Gatto, Phys. Rev. {\bf D81}, 114031(2010).
\bibitem{db3} C.V.Johnson and A.Kundu, JHEP 0812, 053(2011).
\bibitem{db5} V. Skokov, Phys. Rev. {\bf D85}, 034026(2012).
\bibitem{db6} E.S.Fraga, J.Noronha and L.F.Palhares, Phys. Rev. {\bf D87}, 114014(2013).

\bibitem{pnjl1} R.Gatto and M.Ruggieri, Phys. Rev. {\bf D82}, 054027(2010), {\bf D83}, 034016(2011).
\bibitem{pnjl2} M.Ferreira, P.Costa and C.Provid$\hat e$ncia, Phys. Rev. {\bf D89}, 036006(2014).
\bibitem{pnjl3} M.Ferreira, P.Costa, D.P.Menezes, C.Provid$\hat e$ncia and N.N.Scoccola, Phys. Rev. {\bf D89}, 016002(2014).
\bibitem{pnjl4} P.Costa,  M.Ferreira,  H.Hansen, D.P.Menezes and C.Provid$\hat e$ncia, Phys. Rev. {\bf D89}, 056013(2014).
\bibitem{pqm} A.J.Mizher, M.N.Chernodub and E.S.Fraga, Phys. Rev. {\bf D82}, 105016(2010).
\bibitem{mit1} N.O.Agasian and S.M.Fedorov, Phys. Lett {\bf B663}, 445(2008).
\bibitem{mit2} E.S.Fraga and L.F.Palhares, Phys. Rev. {\bf D86}, 016008(2012).

\bibitem{pnjl5} P.N.Meisinger and M.C.Ogilvie, Phys. Lett. {\bf B379}, 163(1996).
\bibitem{pnjl6} P.N.Meisinger, T.R.Miller and M.C.Ogilvie, Phys. Rev. {\bf D65}, 034009(2002).
\bibitem{pnjl7} K.Fukushima, Phys. Lett. {\bf B591}, 277(2004).
\bibitem{pnjl8} A.Mocsy, F.Sannino, and K.Tuominen, Phys. Rev. Lett. {\bf 92}, 182302(2004).
\bibitem{pnjl9} E.Megias, E.Ruiz Arriola, and L.L.Salcedo, Phys. Rev. {\bf D74}, 065005(2006).
\bibitem{pnjl10} C.Ratti, M.A.Thaler, and W.Weise, Phys. Rev. {\bf D73}, 014019(2006).
\bibitem{pnjl11} C.Ratti, M.A.Thaler, and W.Weise, nucl-th/0604025.
\bibitem{pnjl12} S.K.Ghosh {\it et al.}, Phys. Rev. {\bf D73}, 114007(2006).
\bibitem{zhuang} P.F.Zhuang, J.Huefner and S.P.Klevansky, Nucl. Phys. {\bf A576}, 525(1994).
\bibitem{njl1} Y.Nambu and G.Jona-Lasinio, Phys. Rev. {\bf 122}, 345(1961) and {\bf 124}, 246(1961).
\bibitem{njl2} S.P.Klevansky, Rev. Mod. Phys. {\bf 64}, 649(1992).
\bibitem{njl3} M.K.Volkov, Phys. Part. Nucl. {\bf 24}, 35(1993).
\bibitem{njl4} T.Hatsuda and T.Kunihiro, Phys. Rep. {\bf 247}, 221(1994).
\bibitem{njl5} M.Buballa, Phys. Rep. {\bf 407}, 205(2005).
\bibitem{njl6} H.Hansen {\it et al.}, Phys. Rev. {\bf D75}, 065004(2007).
\bibitem{fu}W.J.Fu and Y.X.Liu, Phys. Rev. {\bf D79},074011(2009).
\bibitem{ritus1} Sh.Fayazbakhsh, S.Sadeghian and N.Sadooghi, Phys. Rev. {\bf D86}, 085042(2012).
\bibitem{ritus2} Sh.Fayazbakhsh and N.Sadooghi, Phys. Rev. {\bf D88}, 065030(2013).

%
%


\end{thebibliography}
\end{document}